\documentclass[pre,aps,twocolumn]{revtex4} 
\usepackage{graphicx} 

\usepackage{amsfonts}
\usepackage{pifont}
\usepackage{marvosym}

\newcommand{\flocon}{\mbox{\ding{100}}}

\newcommand{\rhosat}{\rho_{\rm sat}}

\begin{document}

\title{Physical processes causing the formation of penitentes}

\author{P. Claudin}
\author{H. Jarry}
\author{G. Vignoles$^{\mbox{\Mundus}}$}
\author{M. Plapp$^{\flocon}$}
\author{B. Andreotti}
\affiliation{Laboratoire de Physique et M\'ecanique des Milieux H\'et\'erog\`enes,\\
PMMH UMR 7636 ESPCI -- CNRS -- Univ. Paris Diderot -- Univ. P. M. Curie, 10 rue Vauquelin, 75005, Paris, France.}
\affiliation{$^{\mbox{\Mundus}}$Laboratoire des Composites ThermoStructuraux,\\
LCTS UMR 5801 CNRS -- UB1 -- CEA -- Safran, 3 all\'ee La Boetie, 33600 Pessac, France.}
\affiliation{$^{\flocon}$Condensed Matter Physics, Ecole Polytechnique, CNRS, 91128 Palaiseau, France.}

\begin{abstract}
Snow penitentes form in sublimation conditions by differential ablation. Here we investigate the physical processes at the initial stage of penitente growth and perform the linear stability analysis of a flat surface submitted to the solar heat flux. We show that these patterns do not simply result from the self-illumination of the surface --a scale-free process-- but are primarily controlled by vapor diffusion and heat conduction. The wavelength at which snow penitentes emerge is derived and discussed. We found that it is controlled by aerodynamic mixing of vapor above the ice surface.
\end{abstract}

\maketitle

\section{Introduction}
\label{Introduction}
\begin{figure}
\includegraphics{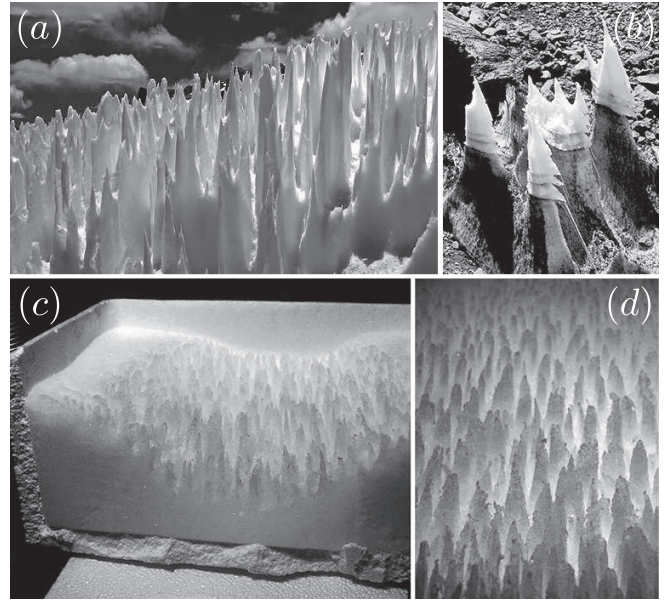}
\caption{(a,b) Photographs of natural penitentes on the Aconcagua mountain (Argentina). Peak separation: a few tens of~cm. Photo credits: Paul Dubuc. (c,d) Micropenitentes in the laboratory, from Ref. \cite{BBB06}. Peak separation $\simeq 1$~cm.}
\label{Fig1}
\end{figure}

Penitentes are natural patterns made of compact snow or ice (Fig.~\ref{Fig1}). They are typically found in mountains at high altitudes \cite{N39,O41,L54,A58,NL97,SdJ04,CP05} where humidity and temperature are low and solar radiation is intense --penitentes are also expected to form on other planetary bodies \cite{H13}. In these conditions, solid water sublimates when heated, and tall thin spikes oriented toward the main direction of the sun emerge by differential ablation. They have been reproduced at a centimeter scale in laboratory experiments \cite{BBB06}. It has also been argued that conical spikes obtained by irradiation of silicon surfaces with laser pulses are the equivalent of penitentes at a micrometer scale \cite{HFWDM98,PHZFCCZ13}. Melting conditions rather generate ablation hollows on snowfields \cite{RAW87,B01,TMLBN06,MT10}. Their shape is that of shallow cups with sharp edges and are similar to ablation patterns on the surface of meteorites (regmaglypts) \cite{LQ87} and to ripples generated by ion erosion of sputtering targets \cite{M-GCC06}.

It has been suggested that penitentes result from an instability due to a geometrical effect: troughs receive more radiation than crests because of photons diffused by the snow surface \cite{L54,B01,CAMcA14}. As more radiation leads to an enhanced sublimation rate, this effect constitutes a positive feedback mechanism amplifying an undulated topography. However, as this process is scale-free, it immediately raises several questions. With this dynamical mechanism only, how do we explain the selection of the penitente wavelength observed in natural \cite{L54,NL97} and laboratory \cite{BBB06} conditions (Fig.~\ref{Fig1})? What are the mechanisms stabilizing the long wavelengths, and how do we explain such a selection? Is this geometrical effect the only instability mechanism?

A simple model has been proposed by Betterton \cite{B01}, where the growth of penitentes due to self-illumination is balanced by an effective diffusion of the surface height. This diffusion provides a small-scale cut off, but the associated mechanism is not clear. In this paper, we revisit the linear stability analysis of the problem and put emphasis on two specific aspects. First, the light does not directly lead to sublimation. It is absorbed by snow, which is heated, leading to a temperature gradient \emph{toward the interface}. Heat is then transported toward the surface by conduction, from the inside. This is exactly the condition needed for a Mullins-Sekerka type of instability \cite{MS64} to take place, as studied in the context of directional solidification \cite{CCR86} and for pattern formation (e.g. dendrites) in crystal growth \cite{L80}. Our second point is that the sublimation rate depends on the vapor concentration close to the surface, and thus that the evacuation of this vapor away from it plays, through a typical associated length scale $\ell$, an essential role in the dynamics of this instability.

In section~2, we set the starting equations for the modeling of the diffusion of vapor, temperature and light, as well as the expressions of the sublimation rate. We then compute the corresponding base state (section~3) and the surface illumination on a modulated surface due to light reflection (section~4). The linear stability analysis is performed in section~5, and its results are discussed in section~6.

\begin{figure}
\includegraphics{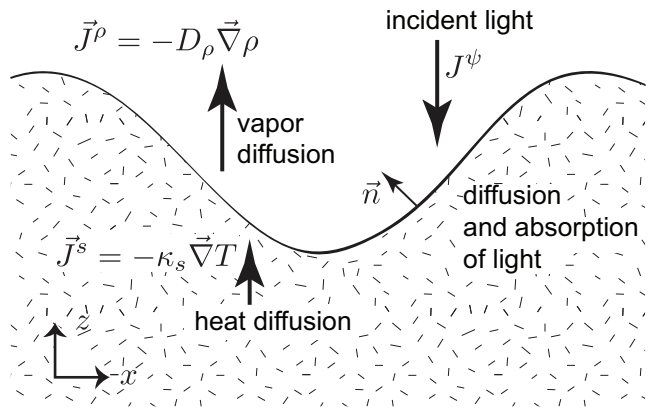}
\caption{Schematic of the system. $z$ is the direction pointing toward the light source. The $x$-axis is perpendicular to it. $\vec n$ denotes the unit vector normal to the interface between the ice block and the air, pointing towards the gas. The bold arrows represent the vapor density, heat and light power fluxes.}
\label{Fig2}
\end{figure}

\section{Model equations}
\label{ModelEquations}
We consider a semi-infinite block of ice submitted to incident light, as schematized in fig.~\ref{Fig2}. Its sublimation is governed by bulk diffusion of temperature, concentration and light, and by conservation laws at the interface.

\subsection{Diffusion of vapor, temperature and light}
Neglecting possible hydrodynamical flows, the evolution of the vapor density $\rho$ (mass of water vapor per unit volume) is governed by a diffusion equation:
\begin{equation}
\partial_t \rho=-\vec \nabla \cdot \vec J^\rho=D_\rho  \vec \nabla^2 \rho.
\label{EquaC}
\end{equation}
$\vec J^\rho=-D_\rho \vec \nabla \rho$ is the diffusive mass flux of water, where $D_\rho$ is the diffusion coefficient of vapor in the air. Its typical value in ambient conditions is $D_\rho \simeq 3 \, 10^{-5}$~m$^2$/s \cite{LW54}. This description is valid at a scale larger than the mean free path of water molecules in air. Similarly, we consider the diffusion of temperature $T$ in the ice
\begin{equation}
\rho_s c_s \partial_t T=-\vec \nabla \cdot \vec J^s+ \psi=\kappa_s  \vec \nabla^2 T+\psi.
\label{EquaTs}
\end{equation}
We neglect it in the gas. $\rho_s \simeq 10^3$~kg/m$^3$ is the ice density, and $c_s \simeq 2 \, 10^3$~J/kg/K is the ice specific heat. $\vec J^s=-\kappa_s \vec \nabla T$ is the heat flux in the solid, where $\kappa_s$ is the ice thermal conductivity. Its typical value is $\kappa_s \simeq 2$~W/m/K, corresponding to an ice thermal diffusivity $\kappa_s/(\rho_s c_s) \simeq 10^{-6}$~m$^2$/s \cite{J68}. The power $\psi$ per unit volume arises from the absorption of the light energy in the ice.

\begin{figure}
\includegraphics{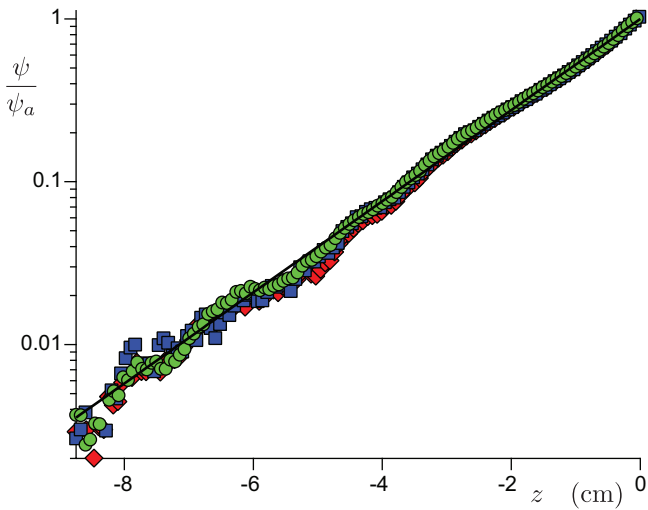}
\caption{(Color online) Decay of light intensity in artificial snow, for blue, green and red wavelengths. The best fit by an exponential (black solid line) gives a decay length $\Lambda \simeq 1.6$~cm. The snow is prepared by spraying ten micron-scale droplets into a flat reservoir of liquid nitrogen. It is shaped into a cubic isotherm $25\times25\times25\,{\rm cm}^3$ box. The snow surface is illuminated with white parallel light using a slide projector. A picture is taken from the side using a calibrated color digital camera. The signals received in the red, blue and green photosensors are averaged over the direction transverse to the light.}
\label{Fig3}
\end{figure}

In the purely diffusive limit (\emph{i.e.} when the absorption coefficient is small with respect to the scattering coefficient), the direction of the light does not have any influence. The light diffusion/absorption equation, described in steady state and in the absence of internal sources, governs the space distribution of the light power per unit area $\varphi$ (or fluence rate, in units of W/m$^{2}$), which takes the form $\Lambda^2 \vec \nabla^2 \varphi-\varphi=0$ \cite{BH83}. $\Lambda$ is a characteristic attenuation length which can be expressed as a function of absorption and scattering coefficients. The value of $\Lambda$ is on the order of a few cm in compacted snow and can be larger in clean ice \cite{WBG06}. We have also measured this length in artificial snow by illuminating a cube of snow with parallel light and taking a lateral picture. We have obtained $\Lambda \simeq 1.6$~cm for this particular snow sample, independent of the light wavelength (Fig.~\ref{Fig3}). The absorbed power per unit volume $\psi$ is proportional to $\varphi$ and is therefore controlled by the same equation:
\begin{equation}
\Lambda^2  \vec \nabla^2 \psi-\psi=0.
\label{EquaLight}
\end{equation}
The surface of the ice is submitted to an insolation corresponding to a light power flux $J^{\psi}$. On Earth, its typical value due to direct sun illumination is $J^{\psi}_0 \simeq 200$~W/m$^{2}$. We denote the albedo as $\omega$ (it varies from $0.2$ to $0.8$ for ice and snow), so that the boundary condition for the absorbed volumetric power $\psi$ is
\begin{equation}
\Lambda^2 \vec \nabla \psi \cdot \vec n= \left(1 - \omega\right) J^\psi ,
\label{BCpsi}
\end{equation}
where $\vec n$ the unit vector normal to the surface, oriented from the solid toward the gas (Fig.~\ref{Fig2}).

\subsection{Sublimation rate} 
The time evolution of the surface elevation $h(x,t)$ is governed by the sublimation rate $q$:
\begin{equation}
\partial_t h=\frac{q}{\vec e_z\cdot \vec n} \, .
\end{equation}
Note that $q$ is negative, as the pattern emerges by progressive ablation of the solid. It obeys three equations simultaneously. The conservation of mass gives:
\begin{equation}
q=- \frac{\vec J^\rho \cdot \vec n}{\rho_s}.
\label{defqMass}
\end{equation}
The conservation of energy gives:
\begin{equation}
q=-\frac{\vec J^s\cdot \vec n}{{\rho_s \mathcal L}},
\label{defqEnergy}
\end{equation}
where ${\mathcal L} \simeq 3 \, 10^6$~J/kg is the sublimation latent heat of the ice. Finally, the dissolution/precipitation kinetics depends on the difference between the actual vapor density at the interface $\rho^i$ and its saturation value $\rhosat$. This gives:
\begin{equation}
q=\alpha \left( \frac{\rho^i-\rhosat(T^i)}{\rho_s}\right).
\label{defqKinetics}
\end{equation}
In this expression, $\alpha$ is a velocity scale, proportional to the characteristic thermal velocity of particles in a gas, $\sqrt{k_B T /m}$, times a desorption probability. We estimate that the value of $\alpha$ lies between $1$ and $100$~m/s. For the sake of simplicity, we neglect its variations with temperature. The saturation density $\rhosat$ is a calibrated function of the temperature \cite{MK05}, here evaluated at the interface $T^i$. We can expand it around the reference temperature $T_0$ as
\begin{equation}
\rhosat \left( T^i \right)-\rhosat \left( T_0 \right) = \rhosat' \left(T^0\right) \left( T^i - T_0 \right),
\label{rhosatExpanded}
\end{equation}
where the prime means the derivative with respect to the temperature. Using the perfect gas law $p/\rho=RT/M_s$, where $R \simeq 8.3$~J/mol/K is the perfect gas constant and $M_s=18 \, 10^{-3}$~kg/mol the molecular weight of water, we can express $\rho' = \frac{\rho}{T} \left( \frac{p' M_s}{\rho R} - 1 \right)$. Now using the Clausius-Clapeyron relationship $p' = \mathcal{L} \rho/T$, we can write
\begin{equation}
\rhosat' (T_0)= \frac{\rhosat (T_0)}{T_0} \left( \frac{M_s \mathcal{L}}{RT_0} - 1 \right).
\label{rhoprime}
\end{equation}
The dimensionless factor in parentheses is on the order of $20$. For a vapor pressure at saturation $p_{\rm sat} \simeq 0.6\, 10^3$~Pa around $273$~K \cite{MK05}, we obtain $\rhosat' \simeq 4 \, 10^{-4}$~kg/m$^3$/K.

Note that in writing down Eq.~(\ref{defqKinetics}) we have neglected the effect of capillarity: the saturation pressure should also depend on the curvature of the interface (local equilibrium, described by the Kelvin equation). In the Mullins-Sekerka analysis \cite{MS64}, the interplay between capillarity and diffusion selects the characteristic length scale of the interfacial instability, but this scale is typically in the micron range. On the much larger length scales of interest here, capillarity can safely be neglected.

\section{Base state}
\label{BaseState}
In order to compute the base state of the problem, we consider that all processes are much faster than the time scale over which the ice surface elevation evolves. The computation is performed in the frame of reference of the surface, which moves downward with respect to the solid ground underneath. The temperature and density fields are therefore stationary. The temperature in the gas is noted $T_0$, and it is also that of the interface:
\begin{equation}
T_0^i =T_0,
\end{equation}
The light volumetric power $\psi$ vanishes asymptotically as $z\to -\infty$ so that  the base state for the light is:
\begin{equation}
\psi_0 = \psi_a e^{z/\Lambda},
\end{equation}
where $\psi_a$ is the interfacial value of $\psi$. With the condition (\ref{BCpsi}), it gives $\psi_a = (1-\omega) J^\psi_0 / \Lambda$. We checked this relation experimentally over three decades (Fig.~\ref{Fig3}). We assume that the thermal flux $J^s_\infty$ vanishes in the bulk of the solid as $z\to -\infty$, far from the surface. The temperature in the solid obeys the equation $\kappa_s  \vec \nabla^2 T+\psi = 0$. The solution is the sum of $-\Lambda^2/\kappa_s \psi$ plus a homogeneous solution ($ \vec \nabla^2 T=0$), which is here simply a constant as $J^s_\infty \to 0$. The base states for the temperature and the flux $\vec{J}^s = -\kappa_s \vec{\nabla} T$ then read:
\begin{eqnarray}
T^s_0 &=& T_0+\frac{1}{\kappa_s}\, \psi_a \Lambda^2 \left(1-e^{z/\Lambda} \right),
\label{Ts0II}\\
J^s_0 &=& \psi_a\Lambda e^{z/\Lambda}.
\label{Js0II}
\end{eqnarray}
The temperature deep inside the solid is therefore larger than in surface and tends to $T_0+\frac{1}{\kappa_s}\, \psi_a \Lambda^2$.

The sublimation rate, as defined in (\ref{defqEnergy}), is the ratio between the heat flux at the interface and the latent heat. We obtain:
\begin{equation}
q_0 = - \frac{\psi_a \Lambda}{\rho_s\mathcal{L}} \, .
\end{equation}
Here, the light power $\psi_a$ is imposed, and the flux of vapor $J_0^\rho$ must adjust following (\ref{defqMass}) to ensure a steady state: $J_0^\rho = - \rho_s q_0$. The density profile reads:
\begin{equation}
\rho_0 = \rho_0^i - \frac{J_0^\rho z}{D_\rho} \, ,
\end{equation}
where $\rho_0^i$ is the vapor density at the interface. For a given temperature $T_0$, $\rho_0^i$ adjusts following the kinetic condition (\ref{defqKinetics}): $\rho_s q_0 = \alpha [ \rho_0^i - \rhosat(T_0) ]$.

\section{Surface illumination on a modulated surface}
\label{SurfaceIllumination}
\begin{figure}
\includegraphics{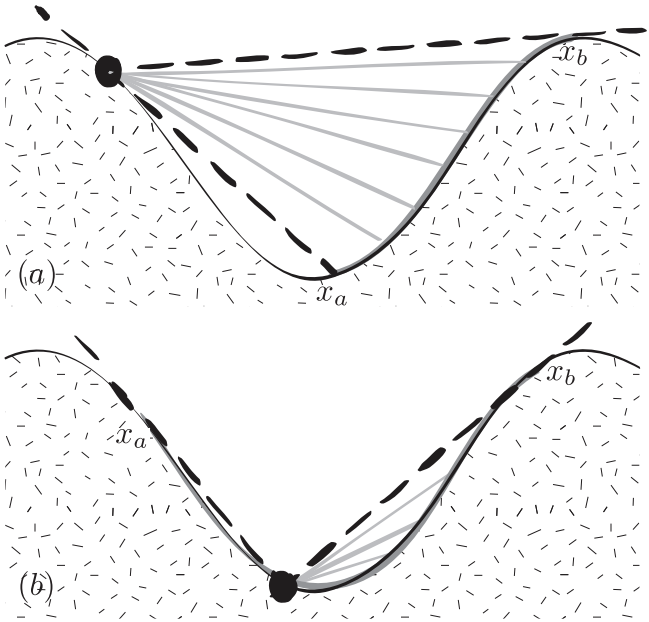}
\caption{A given point $x$ receives light from a portion of the surface. The rays determining the limits of this portion are either tangential to the surface at the point considered (a) or at the point of emission (b). These conditions determine $x_a$ and $x_b$ (Eqs. \ref{bounds1}-\ref{bounds3}).}
\label{Fig5}
\end{figure}

In this section, we determine the illumination of a modulated surface $z=h(x)$ in a way similar to \cite{B01}. Due to a finite albedo $\omega$, a unit surface re-emits a light power flux $\omega J^\psi$, proportional to the power received $J^\psi$. When the interface is flat, none of the re-emitted photons reach the surface again. However, when the surface is modulated, its illumination is partly due to these photons. Assuming isotropy of the re-emission, and a one-dimensional profile, one obtains:
\begin{equation}
J^\psi(x,h)  = J^\psi_0 + \omega \int_{x_a}^{x_b} \frac{|\mathcal{S}|}{\pi} J^\psi(\xi,h(\xi)) \, d\xi \, ,
\label{reemission}
\end{equation}
where $\mathcal{S} d\xi$ is the solid angle through which the element $d\xi$ at position $\xi$ is seen from point $x$, which reads
\begin{equation}
\mathcal{S}(\xi) = \frac{1}{x-\xi} \left [ h'(x) - \frac{h(x)-h(\xi)}{x-\xi} \right ].
\label{mathcalS}
\end{equation}
In this expression, $h'$ is the derivative of the interface profile. The bounds $x_a$ and $x_b$ of this integral both depend on $x$ too. They correspond to positions beyond which the interface cannot be seen from position $x$, due to shadowing (Fig.~\ref{Fig5}).

\begin{figure}
\includegraphics{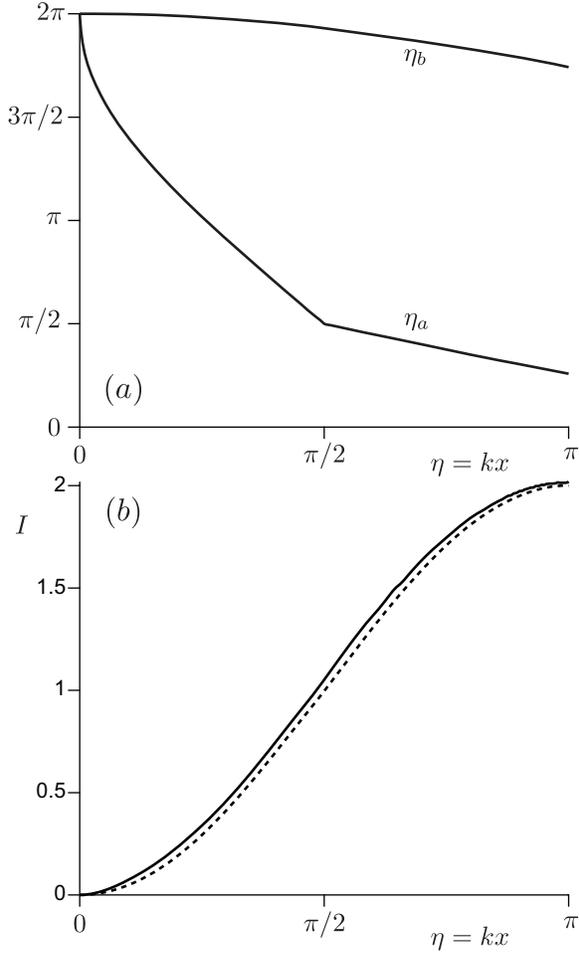}
\caption{(a) Bounds of the integral (\ref{inti}) given by Eqs.~\ref{bounds1}-\ref{bounds3}. (b) Integral $I(\eta)$ (solid line) giving the illumination profile for a sinusoidal surface, compared to the function $1-\cos(kx)$ (dotted line).}
\label{Fig8}
\end{figure}

\begin{figure}
\includegraphics{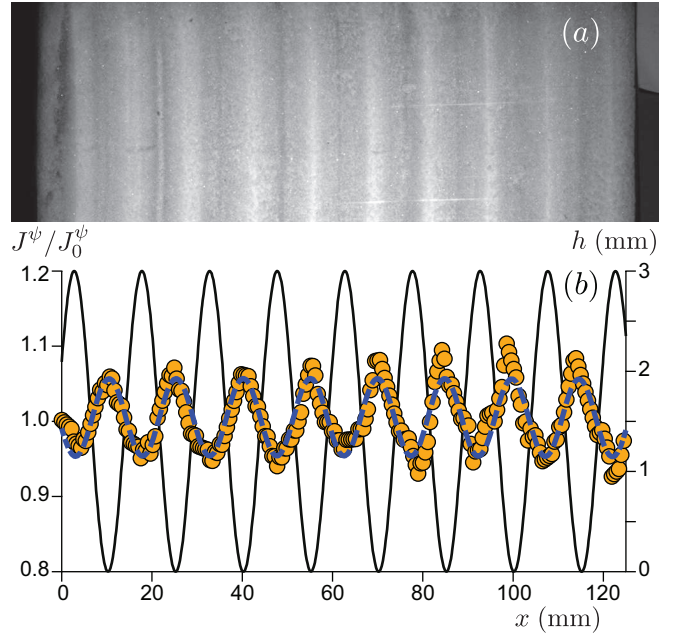}
\caption{(Color online) Surface illumination of a modulated snow surface. The photograph on the top (a) shows the snow block when illuminated from above. It is scaled to correspond to the longitudinal profiles below. Snow is prepared with a sinusoidal elevation profile $h$ of wavelength $\lambda=1.5$~cm and of amplitude $1.5$~mm crest to crest (black solid line). The illumination, rescaled by its average, is measured by means of image analysis and its profile is displayed with orange circles (b). Fitting these data by a sinusoidal function (blue dashed line), one observes an out-of-phase modulation from which one extracts $\Omega \simeq 0.08$ (Eq.~\ref{psiillumination}), which corresponds to $\omega \simeq 0.25$.}
\label{Fig4}
\end{figure}

As the reference state considered is homogeneous, at the linear order, eigen-modes of the illumination operator (\ref{reemission}) are periodic. However, due to the non-local nature of $\mathcal{S}$, they are not Fourier modes, as known in the general context of Fredholm equations. In particular, the illumination of a sinusoidal profile $h(x)=h_1 \cos (kx)$, such as the one shown in Fig.\,\ref{Fig4} is not strictly sinusoidal. For such a function, the contribution to the integral term giving the illumination at first order in $kh_1$ reads:
\begin{equation}
I(\eta)=\int_{\eta_a}^{\eta_b} \left | \sin\eta + \frac{\cos\eta-\cos\eta'}{\eta-\eta'} \right |\frac{d\eta'}{|\eta-\eta'|} \, ,
\label{inti}
\end{equation}
where $\eta=kx$. The boundaries of the integral, $\eta_a(\eta)$ and $\eta_b(\eta)$, correspond to rays that are tangent to the surface (Fig.~\ref{Fig5}). For $0<\eta<\pi$, they are solutions of
\begin{eqnarray}
\sin\eta + \frac{\cos\eta_a-\cos\eta}{\eta_a-\eta} = 0 \; &\mbox{for}& \; 0 \leq \eta \leq \pi/2 \qquad
\label{bounds1} \\
\sin\eta_a + \frac{\cos\eta-\cos\eta_a}{\eta-\eta_a} = 0 \; &\mbox{for}& \; \pi/2 \leq \eta \leq \pi \qquad
\label{bounds2} \\
\sin\eta_b + \frac{\cos\eta_b-\cos\eta}{\eta_b-\eta} = 0
\label{bounds3}
\end{eqnarray}
For $\pi<\eta<2\pi$, the bounds are obtained by symmetry. The dependence of these bounds on $\eta$ is displayed in Fig.~\ref{Fig8}a. However, the non-harmonic contribution of the modes turn out to be negligible and the integral (\ref{inti}) is numerically found to be very close to the function $1-\cos(kx)$ (Fig.~\ref{Fig8}b). For the linear stability performed here, the light volumetric power at the interface can be approximately written as
\begin{equation}
J^\psi(x) = J^\psi_0 \left [ 1 + \Omega kh_1 \left (1-\cos(kx) \right ) \right ],
\label{psiillumination}
\end{equation}
where $\Omega= \frac{\omega}{\pi}$.

\section{Linear stability analysis}
\label{LSA}
We perform the linear stability analysis by using a complex variable --the real part of the equations is understood. We consider an undulated interface of the form $h=h_1\exp(ikx+\sigma t)$, where $kh_1 \ll 1$.

\subsection{Light power profile}
The first order correction for the light power density $\psi$ derives from the Laplace equation (\ref{EquaLight}):
\begin{equation}
\psi = \psi_0(z) + \psi_1 e^{\sqrt{k^2+\Lambda^{-2}}z+ikx+\sigma t}.
\label{psi1a}
\end{equation}
The disturbance to the light power flux at the interface can be computed from (\ref{BCpsi}) as:
\begin{equation}
J^\psi_1 = \frac{\Lambda^2 }{1-\omega} \left(\psi_a \frac{h_1}{\Lambda^2}+ \sqrt{k^2+\Lambda^{-2}} \psi_1 \right) e^{ikx+\sigma t}.
\label{Jpsi1a}
\end{equation}
From (\ref{psiillumination}) and neglecting the homogeneous first order term, the light power flux is well approximated by 
\begin{equation}
J^\psi_1 = - J^\psi_0 \Omega k h_1 e^{ikx+\sigma t},
\label{Jpsi}
\end{equation}
from which, in comparison to (\ref{psi1a}) and recalling that $J_0^\psi = \Lambda \psi_a/(1-\omega)$, one can deduce the following expression for $\psi_1$:
\begin{equation}
\psi_1 = - \psi_a k h_1 \frac{ \Omega + \frac{1}{k\Lambda} }{ \sqrt{1+k^2\Lambda^{2}}}.
\label{psi1}
\end{equation}
The first term in this expression encodes the fact that, due to reflections at the surface, troughs are more illuminated than crests.  We can see that the second term adds up: the light power density is also smaller beneath the crests because the light goes through a larger amount of matter.

\subsection{Temperature profile}
The temperature disturbance in ice is composed of two terms: a term that follows the source term proportional to $-\psi$ in Eq.~(\ref{EquaTs}), and a solution of the homogeneous Laplace equation. It then reads:
\begin{eqnarray}
T^s = T^s_0(z) & - & \frac{\Lambda^2}{\kappa_s} \psi_1 e^{\sqrt{k^2+\Lambda^{-2}}z+ikx+\sigma t}
\nonumber \\
& + & T_1^s e^{kz+ikx+\sigma t}.
\label{TsII}
\end{eqnarray}
The relevant temperature is not $T_1^s$ but the interfacial temperature $T^i$. At first order we obtain:
\begin{eqnarray}
T^i_1
&=& - \frac{1}{\kappa_s} \psi_a \Lambda h_1 -  \frac{\Lambda^2}{\kappa_s} \psi_1 + T_1^s \nonumber\\
&=&   \frac{\Lambda  \psi_a  h_1}{\kappa_s} \left(\frac{ \Omega k\Lambda+ 1 }{ \sqrt{1+k^2\Lambda^{2}}}-1 \right) + T_1^s.
\end{eqnarray}
The full temperature field finally reads:
\begin{eqnarray}
T^s &=& T^s_0(z)
\label{TsIIfull} \\
& + & \frac{\Lambda}{\kappa_s} \psi_a  h_1 \frac{ \Omega k\Lambda + 1}{ \sqrt{1+k^2\Lambda^{2}}} e^{\sqrt{k^2+\Lambda^{-2}}z+ikx+\sigma t}
\nonumber \\
& + & \left [T_1^i-   \left(\frac{ \Omega k\Lambda+ 1 }{ \sqrt{1+k^2\Lambda^{2}}}-1 \right) \frac{\Lambda  \psi_a  h_1}{\kappa_s}\right] e^{kz+ikx+\sigma t}.
\nonumber
\end{eqnarray}
The corresponding heat flux in the ice block can be computed from $\vec{J}^s = -\kappa_s \vec{\nabla}T^s$ and its normal component reads:
\begin{eqnarray}
&& \vec{J}^s \cdot \vec{n} = J^s_0(z) 
\label{Jsnormal} \\
&& \quad - \, \psi_a  h_1\left ( \Omega k\Lambda + 1 \right ) e^{\sqrt{k^2+\Lambda^{-2}}z+ikx+\sigma t} 
\nonumber \\
&& \quad - \left [k \kappa_s T_1^i - \left(\frac{ \Omega k\Lambda+ 1 }{ \sqrt{1+k^2\Lambda^{2}}}-1 \right) k\Lambda  \psi_a  h_1 \right] e^{kz+ikx+\sigma t}.
\nonumber 
\end{eqnarray}
Note that, at the first order, the normal unit vector $vec n$ is vertical. Evaluating this expression at the interface, we obtain from (\ref{defqEnergy}) the following expression for the modulation of the sublimation rate:
\begin{equation}
\rho_s\mathcal{L} q_1   =  k \kappa_s T_1^i -\left(\frac{ \Omega k\Lambda+ 1 }{ \sqrt{1+k^2\Lambda^{2}}}-1-\Omega \right)k \Lambda \psi_a h_1.
\label{Jsnormal1}
\end{equation}
%

\subsection{Vapor density profile}
Following (\ref{EquaC}), the vapor density takes the generic form:
\begin{equation}
\rho =  \rho_0(z) + \left( \rho_1^- e^{-kz}+\rho_1^+ e^{kz} \right)e^{ikx+\sigma t}.
\label{defrho1pm}
\end{equation}
With these notations, the density correction at the interface is given by:
\begin{equation}
\rho_1^i = -J_0^\rho h_1/D_\rho + \rho_1^- + \rho_1^+.
\label{defrho1i}
\end{equation}
We consider that there exists a boundary layer of thickness $\ell$, above which air is permanently kept at constant humidity. Introducing this length is a way to abstract aerodynamical processes and to remain general. $\rho$ is therefore imposed at a distance $\ell$ from the ice surface, so that its first order correction at $z=h_1 + \ell$ must vanish:
\begin{equation}
-J_0^\rho h_1/D_\rho + \rho_1^- e^{-k\ell} + \rho_1^+ e^{k\ell} = 0.
\label{ellpossibility1}
\end{equation}
Equations (\ref{defrho1i}) and (\ref{ellpossibility1}) can be solved for $\rho_1^\pm$ and the resulting vapor density profile reads:
\begin{eqnarray}
\rho =  \rho_0(z) & + & \left( \rho_1^i \frac{\sinh[k(\ell-z)]}{\sinh(k\ell)} \right .
\nonumber \\
& + & \left . \frac{J_0^\rho h_1}{D_\rho}\frac{\cosh[k(\ell/2-z)]}{\cosh(k\ell/2)}  \right)e^{ikx+\sigma t}.
\label{densityprofile1}
\end{eqnarray}
The corresponding vapor flux can be computed from $\vec{J}^\rho = -D_\rho \vec{\nabla}\rho$ and its normal component reads:
\begin{eqnarray}
\vec{J}^\rho \cdot \vec{n} = J^\rho_0 & + & k D_\rho \left (\rho_1^i \frac{\cosh[k(\ell-z)]}{\sinh(k\ell)} \right .
\nonumber \\
& + & \left . \frac{J_0^\rho h_1}{D_\rho}\frac{\sinh[k(\ell/2-z)]}{\cosh(k\ell/2)}  \right)e^{ikx+\sigma t}.
\label{Jrhonormal}
\end{eqnarray}
Evaluating this expression at the interface, from (\ref{defqMass}) and recalling that $J_0^\rho=\psi_a \Lambda/ \mathcal{L}$, we obtain the following expression for the modulation of the sublimation rate:
\begin{equation}
\rho_s q_1 = - \frac{D_\rho k \rho_1^i}{\tanh(k\ell)}-\frac{\psi_a \Lambda}{\mathcal{L}} {\tanh\left(\frac{k\ell}2\right)}  k h_1.
\label{Jrhonormal1}
\end{equation}
%

\subsection{Dispersion relation}
As discussed around Eqs. (\ref{defqMass}-\ref{defqKinetics}), the sublimation rate modulation $q_1 = \sigma h_1$ of the pattern, where $\sigma$ is  the growth rate, simultaneously obeys three equations. The conservation of energy, which derives from the heat flux (\ref{Jsnormal1}) evaluated at the interface leads to:
\begin{equation}
\frac{\rho_s \mathcal{L}} {\psi_a}\sigma  = k\Lambda \frac{\kappa_s}{\psi_a \Lambda} \frac{T_1^i}{h_1} + k \Lambda \left(1+\Omega -\frac{ \Omega k\Lambda+ 1 }{ \sqrt{1+k^2\Lambda^{2}}}\right).
\label{sigmaheatflux}
\end{equation}
The conservation of mass, derived from the vapor flux (\ref{Jrhonormal1}) at the interface, reads:
\begin{equation}
\frac{\rho_s \mathcal{L}} {\psi_a}\sigma  = - \frac{k \Lambda}{\tanh(k\ell)} \frac{\mathcal{L} D_\rho}{\psi_a \Lambda} \frac{ \rho_1^i}{h_1} - {\tanh\left(\frac{k\ell}2\right)}  k\Lambda.
\label{sigmavaporflux1}
\end{equation}
Finally, the third equation comes from the kinetics:
\begin{equation}
q_1 =\alpha\frac{ \rho_1^i - \rhosat' T_1^i}{\rho_s} \, ,
\end{equation}
and gives:
\begin{equation}
\frac{\rho_s \mathcal{L}} {\psi_a} \sigma  = \frac{\alpha  \mathcal{L}} {\psi_a}\frac{\rho_1^i}{h_1} - \frac{\rhosat'  \alpha \mathcal{L}}{\psi_a} \frac{T_1^i}{h_1}
\label{sigmakinetics}
\end{equation}

We introduce two dimensionless numbers. ${\mathcal P}$ compares the influence of heat conductivity and mass diffusion:
\begin{equation}
{\mathcal P}=\frac{\kappa_s}{\rhosat' D_\rho \mathcal{L}}
\label{defmathcalP}
\end{equation}
and ${\mathcal R}$ compares the influence of heat conductivity and kinetics
\begin{equation}
{\mathcal R}=\frac{\kappa_s}{\rhosat'  \alpha \Lambda \mathcal{L}}\, .
\label{defmathcalR}
\end{equation}
Assuming the instantaneous equilibrium between the vapor and its saturated value corresponds to $\mathcal{R} \to 0$. With the numerical values of the different parameters given in section~\ref{ModelEquations}, we can estimate these two dimensionless numbers as ${\mathcal P}\simeq 60$ and ${\mathcal R}\lesssim 0.1$.

The main dependence of $\mathcal{P}$ with the temperature comes from the factor $\rho'_{\rm sat}$: it is related to the vapor density, or the vapor pressure, which decreases in an exponential manner when $T$ is lowered \cite{MK05}. Larger $\mathcal{P}$ are thus expected for lower temperatures. For instance, around $250$~K, we have $\rho'_{\rm sat} \simeq 7 \, 10^{-5}$~kg/m$^3$/K, and thus $\mathcal{P} \simeq 300$. Interestingly, neglecting $1$ in front of $M_s \mathcal{L} / (R T_0)$ in (\ref{rhoprime}), one can express the temperature derivative of the vapor density as $\rho'_{\rm sat} \approx \rho_{\rm sat} M_s \mathcal{L} / (R T_0^2)$. This allows us to rewrite $\mathcal{P}$ as the product of three factors:
\begin{equation}
\mathcal{P} \approx \frac{\kappa_s}{\rho_s c_s D_{\rho}} \times \frac{RT_0}{M_s\mathcal{L}} \times \frac{\rho_s c_s T_0}{\rho_{\rm sat}\mathcal{L}}
\label{mathcalPin3factors}
\end{equation}
Besides the competition between diffusive coefficients of heat in the solid and of mass in the gas, one can identify two other quantities: (the inverse of) a dimensionless sublimation heat and a ratio between an internal energy and a sublimation energy. It is interesting to compare these different factors for different materials around their temperature of sublimation. One can take the example of carbon. As a matter of fact, the mechanism that we discuss here could be at the origin of  scallops or cross-hatching that has been evidenced on nose tips made of carbon and placed in high-enthalpy, high velocity plasma flows simulating atmospheric re-entry conditions \cite{HW76}; it constitutes a more plausible scenario than a previous one \cite{DVGA05}, which contained an unnoticed algebraic error. Taking a typical temperature $T_0 \simeq 3800$~K, the physical parameters are $\kappa_s \simeq 200$~W/m/K, $\mathcal{L} \simeq 6 \, 10^7$~J/kg, $D_\rho \simeq 5 \, 10^{-4}$~m$^2$/s, $\rho_s \simeq 2 \, 10^3$~kg/m$^3$, $M_s=12 \, 10^{-3}$~kg/mol, $c_s \simeq 2 \, 10^3$~J/kg/K and $p_{\rm sat} \simeq 10^3$~Pa corresponding to $\rho_{\rm sat} \simeq 4 \, 10^{-4}$~kg/m$^3$. Combining these numbers, we obtain: $\kappa_s/(\rho_s c_s D_\rho) \simeq 10^{-1}$, $M_s\mathcal{L}/(RT_0) \simeq 23$ and $\rho_s c_s T_0/(\rho_{\rm sat}\mathcal{L)} \simeq 6 \, 10^5$, whose values are so different to those for ice: respectively $3 \, 10^{-2}$, $24$ and $4 \, 10^4$. The parameter $\mathcal{P}$ for carbon is eventually around $2600$, an order of magnitude larger than for ice.

Finally, the three above equations (\ref{sigmaheatflux}, \ref{sigmavaporflux1}, \ref{sigmakinetics}) can be combined to give the following dispersion relation relating the growth rate to the wavenumber:
\begin{eqnarray}
\frac{\rho_s \mathcal{L}} {\psi_a} \sigma 
& = & \frac{k\Lambda }{1+{\mathcal P} \tanh(k \ell)+{\mathcal R} k\Lambda}
\nonumber \\
& \times & \left [ \left( 1-\frac{1}{\sqrt{1+k^2\Lambda^2}} \right) +  \Omega \left( 1- \frac{k \Lambda}{\sqrt{1+k^2\Lambda^2}}\right) \right .
\nonumber \\
& - & \left . \left(1 - \frac{1}{\cosh(k \ell)} \right) {\mathcal P} \, \right ].
\label{DispersionRelation1}
\end{eqnarray}
The two first terms in the square brackets are positive and respectively correspond to the destabilizing role of the inverted temperature gradient and the self-illuminating process, proportional to $\Omega$. The third contribution is negative, coming from the stabilizing effect of vapor diffusion above the surface.

\begin{figure}
\includegraphics{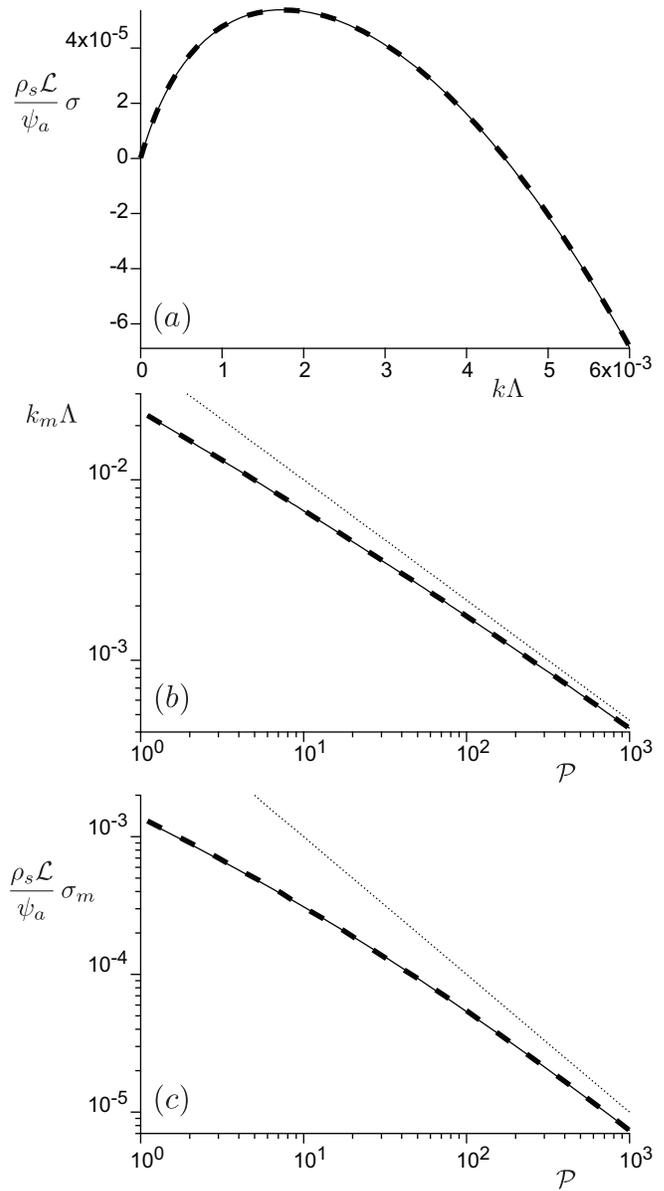}
\caption{(a) Dimensionless growth rate as a function of $k\Lambda$ for $\mathcal{P}=100$, $\ell/\Lambda=10$, $\Omega=0.1$ and $\mathcal{R}=0$. Solid line: full dispersion relation (\ref{DispersionRelation1}). Dashed line: approximation (\ref{DispersionRelation5}). (b) Most unstable wavenumber $k_m \Lambda$ as a function of $\mathcal{P}$, all other parameters kept the same as in (a). Solid line: numerical computation from the full dispersion relation (\ref{DispersionRelation1}). Dashed line: $k_m \Lambda$ deduced from (\ref{km}, \ref{grandA}). Dotted line: $k_m \Lambda$ computed from the asymptotic scaling (\ref{kmsimple}). (c) Same as (b) for the growth rate $\sigma_m$ of the most unstable mode.}
\label{FigRelDispCase1}
\end{figure}

\section{Discussion}
\label{Discussion}

We can now study and discuss the different regimes of the dispersion relation (\ref{DispersionRelation1}).
We recall that we have not included in our calculation the effect of surface tension. As discussed previously, local equilibrium at the interface provides a stabilizing effect that acts on small length scales. Therefore, taking into account capillarity would alter the dispersion relation in the limit of large $k$. Nevertheless, we proceed by discussing the dispersion relation that we have derived above, in order to clarify the interplay of illumination, heat conduction and vapor diffusion in the interfacial instability. As will be seen, the typical relevant length scales that are found in this analysis are large enough to neglect capillarity.

\subsection{Simple and large-$k$ limits}
The problem simplifies in the limit where the whole illumination power is used for sublimation. This corresponds to the triple limit $ \mathcal P \to 0$, $\mathcal R \to 0$ and $k\Lambda \ll 1$:
\begin{equation}
\sigma   =  \frac{\psi_a}{\rho_s \mathcal{L}} \Omega k \Lambda = \frac{(1-\omega) J^\psi_0}{\rho_s \mathcal{L}} \Omega k.
\end{equation}
This expression shows that the growth rate is unconditionally positive and proportional to the albedo and to the wavenumber. Accounting for a finite penetration length $\Lambda$, but keeping the limit $ \mathcal P \to 0$ and $\mathcal R \to 0$, the dispersion relation becomes:
\begin{equation}
\sigma   = \frac{\psi_a}{\rho_s \mathcal{L}} k \Lambda \left [ 1-\frac{1}{\sqrt{1+k^2\Lambda^2}} +  \Omega \left( 1- \frac{k \Lambda}{\sqrt{1+k^2\Lambda^2}}\right)\right].
\label{DispersionRelation2}
\end{equation}
One can immediately see that the growth rate is still unconditionally positive and diverges when $k \to \infty$. This means that there is no wavelength selection: arbitrarily small scale structures can emerge. Betterton \cite{B01} fixed this problem by introducing a phenomenological diffusive term to encode in a simple form the processes leading to a small-scale cutoff. Here, we can see the role played by the parameter $\mathcal{P}$ in relation to the diffusion of the vapor above the interface: the large-$k$ limit of (\ref{DispersionRelation1}) shows either a linear asymptotic behavior $\sigma \sim \frac{\psi_a}{\rho_s \mathcal{L}} k \Lambda \frac{1-\mathcal{P}}{1+\mathcal{P}}$ when $\mathcal{R}=0$, or a growth rate that tends to a constant $\sigma \sim \frac{\psi_a}{\rho_s \mathcal{L}} \frac{1-\mathcal{P}}{\mathcal{R}}$ for a non-vanishing $\mathcal{R}$. In both cases, the growth rate keeps positive at small scales when $\mathcal{P}<1$. Conversely, large wavenumbers are stable when $\mathcal{P}>1$ and a wavelength selection is possible. We shall work under this assumption in the following analysis.

\begin{figure}
\includegraphics{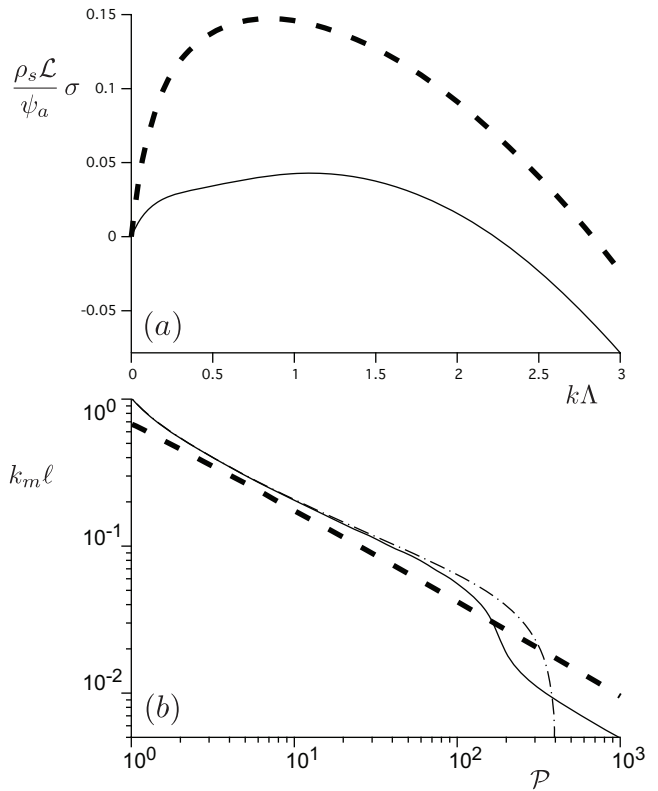}
\caption{(a) Dimensionless growth rate as a function of $k\Lambda$ for $\mathcal{P}=100$, $\ell/\Lambda=0.05$, $\Omega=0.3$ and $\mathcal{R}=0$. Solid line: full dispersion relation (\ref{DispersionRelation1}). Dashed line: approximation (\ref{DispersionRelation6}). (b) Most unstable wavenumber $k_m \ell$ as a function of $\mathcal{P}$. Solid line: numerical computation from the full dispersion relation (\ref{DispersionRelation1}) with all other parameters kept the same as in (a). Dotted-dashed line: idem but with $\Omega=0$. In this case $k_m \ell$ vanishes when $\mathcal{P} \to (\Lambda/\ell)^2 = 400$, like a square root (Eq.~\ref{kmOmega0Pmax}). Dashed line: most unstable wavenumber corresponding to the approximation (\ref{DispersionRelation6}), i.e. $k_m \ell$ deduced from (\ref{km}), where $\Omega=1$ is set in (\ref{grandA}).}
\label{FigRelDispCase3}
\end{figure}

\subsection{Analysis of the dispersion relation for $\ell/\Lambda>1$}
We assume for simplicity that sublimation is not limited by the kinetics ($\mathcal{R} = 0$). We furthermore consider the limit where the absorption length $\Lambda$ is smaller than all other lengthscales ($\ell/\Lambda>1$). The numerical investigation of (\ref{DispersionRelation1}) shows that its behavior can be analyzed in the regime of small $k \Lambda$ and small $k \ell$, for which the growth rate can be approximated as
\begin{equation}
\frac{\rho_s \mathcal{L}} {\psi_a} \sigma  = \frac{k\Lambda}{1+\mathcal{P}k\ell} \left[ \Omega - \frac{1}{2} (k \ell)^2 \mathcal{P} \right].
\label{DispersionRelation5}
\end{equation}
As illustrated in Fig.~\ref{FigRelDispCase1}a, this expression is indeed a very good approximation of the full dispersion relation (\ref{DispersionRelation1}), which shows an unstable ($\sigma>0$) range at small wavenumbers whereas large $k$ are stable ($\sigma<0$). The most unstable wavenumber $k_m$ corresponding to the above expression verifies $(\mathcal{P}k_m \ell)^3 + \frac{3}{2}(\mathcal{P}k_m \ell)^2 - \Omega\mathcal{P}=0$, whose solution is
\begin{equation}
k_m \ell = \frac{1}{2\mathcal{P}} \left( A^{1/3} + \frac{1}{A^{1/3}} - 1 \right),
\label{km}
\end{equation}
with
\begin{equation}
A = 4 \Omega\mathcal{P} - 2\sqrt{2}\sqrt{2(\Omega\mathcal{P})^2 - \Omega\mathcal{P}} - 1.
\label{grandA}
\end{equation}
Note that this expression is valid for $\Omega\mathcal{P}>1/2$ only, otherwise a more complicated formula for $k_m \ell$ applies. For $\Omega \mathcal{P} \gg 1$, (\ref{km}) simplifies into
\begin{equation}
k_m \ell \sim \frac{\Omega^{1/3}}{\mathcal{P}^{2/3}} \, .
\label{kmsimple}
\end{equation}
As shown in Fig.~\ref{FigRelDispCase1}b, this scaling law is asymptotically verified by the numerical computation of $k_m$ from the full equation (\ref{DispersionRelation1}), of which expression (\ref{km}) is almost a perfect approximation. Similarly, the corresponding growth rate asymptotically scales as
\begin{equation}
\frac{\rho_s \mathcal{L}}{\psi_a} \sigma_m \sim \frac{\Lambda}{\ell} \, \frac{\Omega}{\mathcal{P}} ,
\label{sigmamsimple}
\end{equation}
as shown in Fig.~\ref{FigRelDispCase1}c.

When neglecting self-illumination ($\Omega = 0$), the expansion of (\ref{DispersionRelation1}) at small $k\Lambda$ shows that the leading term is cubic: $\frac{\rho_s \mathcal{L}} {\psi_a} \sigma \sim \frac{1}{2} \left[ 1 - \left( \frac{\ell}{\Lambda} \right)^2 \mathcal{P} \right] (k \Lambda)^3$. Recalling that $\mathcal{P}>1$ is required to ensure a non-diverging large-$k$ behavior, this term is negative for $\ell/\Lambda>1$. The interface is then always stable in this case without any effect of the self-illumination process.

\begin{figure}
\includegraphics{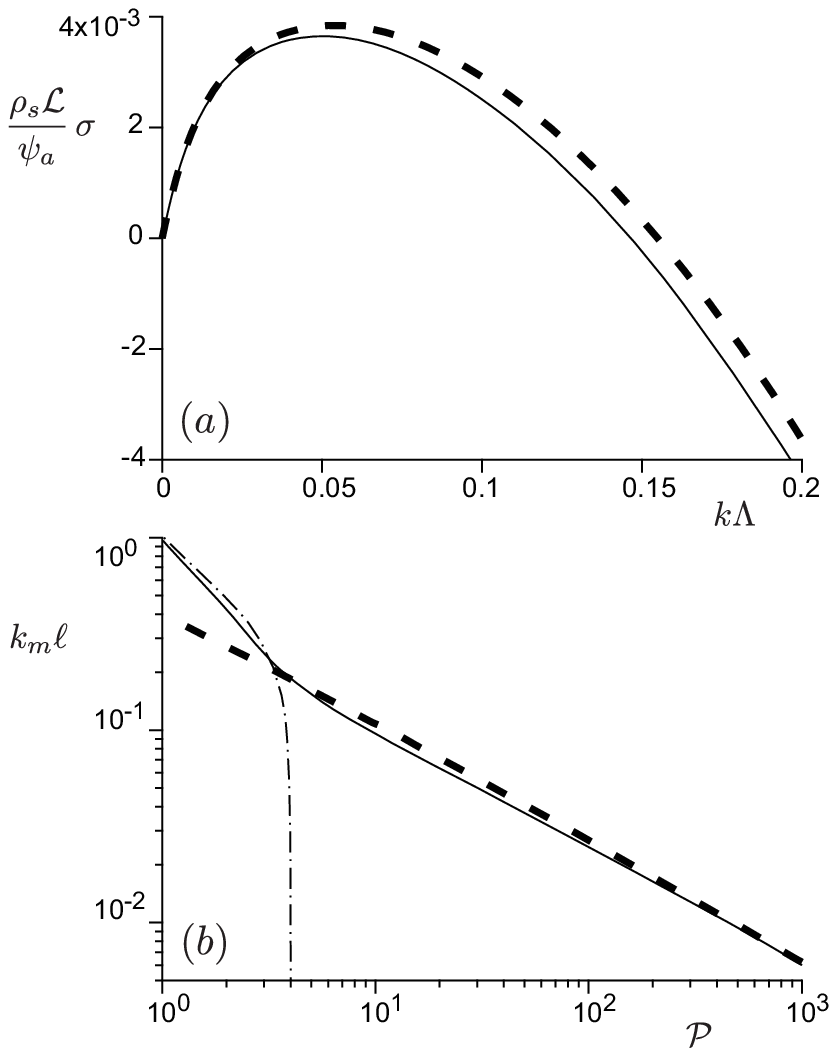}
\caption{(a) Dimensionless growth rate as a function of $k\Lambda$ for $\mathcal{P}=100$, $\ell/\Lambda=0.5$, $\Omega=0.3$ and $\mathcal{R}=0$. Solid line: full dispersion relation (\ref{DispersionRelation1}). Dashed line: approximation (\ref{DispersionRelation5}). (b) Most unstable wavenumber $k_m \ell$ as a function of $\mathcal{P}$. Solid line: numerical computation from the full dispersion relation (\ref{DispersionRelation1}) with all other parameters kept the same as in (a). Dotted-dashed line: idem but with $\Omega=0$. Dashed line: most unstable wavenumber corresponding to the approximation (\ref{DispersionRelation5}), i.e. $k_m \ell$ deduced from (\ref{km},\ref{grandA}).}
\label{FigRelDispCase2}
\end{figure}

\subsection{Analysis of the dispersion relation for $\ell/\Lambda<1$}
Let us first continue with the limit of negligible self-illumination ($\Omega=0$). The system is unstable for a parameter $\mathcal{P}$ in the range $1 < \mathcal{P} < \left( \Lambda/\ell \right)^2$. For $\mathcal{P} \to \left( \Lambda/\ell \right)^2$, the growth rate can be approximated by expanding (\ref{DispersionRelation1}) at small $k\Lambda$ up to the fifth order:
\begin{eqnarray}
\frac{\rho_s \mathcal{L}} {\psi_a} \sigma & \sim & \frac{1}{2} \left[ 1 - \left( \frac{\ell}{\Lambda} \right)^2 \mathcal{P} \right] \left[ (k \Lambda)^3 - \left( \frac{\ell}{\Lambda} \right) \mathcal{P} (k \Lambda)^4 \right]
\nonumber \\
& - & \frac{1}{24} \left[ 9 - 5 \left( \frac{\ell}{\Lambda} \right)^4 \mathcal{P}  \right] (k \Lambda)^5,
\label{RelDispExpanded}
\end{eqnarray}
leading to a most unstable wavenumber vanishing as
\begin{equation}
k_m \Lambda \sim \sqrt{\frac{4}{5} \left[ 1 - \left( \frac{\ell}{\Lambda} \right)^2 \mathcal{P} \right]}
\label{kmOmega0Pmax}
\end{equation}
in the limit of large enough $\Lambda/\ell$. For the parameter $\mathcal{P}$ small enough with respect to its upper bound, $k_m \Lambda$ is found to be of order one or larger. Neglecting the term $1/\sqrt{1+k^2\Lambda^2}$ in front of $1$, and assuming that the condition $k\ell \ll 1$ is still valid, the dispersion relation (\ref{DispersionRelation1}) can be approximated as
\begin{equation}
\frac{\rho_s \mathcal{L}} {\psi_a} \sigma  = \frac{k\Lambda}{1+\mathcal{P}k\ell} \left[ 1 - \frac{1}{2} (k \ell)^2 \mathcal{P} \right].
\label{DispersionRelation6}
\end{equation}
This expression resembles (\ref{DispersionRelation5}), and the corresponding most unstable mode is then identical to (\ref{km}), but where one should formally set $\Omega=1$ in the expression (\ref{grandA}) for $A$ -- recall we are discussing the case $\Omega=0$. In fact, in the limit $\mathcal{P} \to 1$, one can show that $\frac{\rho_s \mathcal{L}} {\psi_a} \sigma \sim k \Lambda e^{-k \ell}$ in the regime $k \Lambda \gg 1$, which gives $k_m \ell =1$. The assumption that $k \ell \ll 1$ to derive (\ref{DispersionRelation6}) is therefore partly valid only.

Now considering a finite albedo ($\Omega > 0$), the behavior of the dispersion relation is not affected at small enough $\mathcal{P}$. The reason is that, in (\ref{DispersionRelation1}), the factor of $\Omega$ vanishes at $k \Lambda \gg 1$ and the contribution of the self-illumination can thus be neglected in front of the term related to the inverted temperature gradient. However, this term suppresses the critical behavior of $k_m$ when $\mathcal{P}$ tends to $\left( \Lambda/\ell \right)^2$, and beyond this value one recovers a growth rate dominated by small $k \Lambda$ and $k \ell$, i.e. well described by (\ref{DispersionRelation5}), with $k_m \ell$ given by (\ref{km}, \ref{grandA}).

We illustrate these results in Figs.~\ref{FigRelDispCase3} and \ref{FigRelDispCase2}. One can see that, as expected, the approximation (\ref{DispersionRelation6}) developed for $\mathcal{P} \ll \left( \Lambda/\ell \right)^2$, for which relevant $k \Lambda$ are larger than unity, is rough (Fig.~\ref{FigRelDispCase3}a). However, the description of the decrease of $k_m \ell$ with $\mathcal{P}$ is still qualitatively correct (Fig.~\ref{FigRelDispCase3}b). On the opposite, for $\mathcal{P} \gg \left( \Lambda/\ell \right)^2$, which makes sense for finite $\Omega$ only, the approximation (\ref{DispersionRelation5}) of the dispersion relation is good (\ref{FigRelDispCase2}a), and the corresponding prediction of the most unstable wavenumber is quantitative (\ref{FigRelDispCase2}b).

\subsection{Effect of a finite $\mathcal{R}$}
Let us now investigate finite values of $\mathcal{R}$. Because this number exclusively enters the dispersion relation as the third term of the denominator of (\ref{DispersionRelation1}), it can only have a significant effect when $\mathcal{R} k\Lambda$ is larger than $1 + \mathcal{P} k \ell$. In the case $\ell/\Lambda \gg1$, this can only occur if $\mathcal{R}$ is much larger than $\mathcal{P}$, which is unlikely given the expected values of these numbers discussed in the previous section. In the case $\ell/\Lambda \ll 1$, large enough values of $\mathcal{R}$ induce a slight decrease of $k_m$, quantitatively similar to an increase of $\mathcal{P}$. In conclusion, no qualitative difference is expected with finite values of $\mathcal{R}$ in comparison to the results described above.

\begin{figure}
\includegraphics{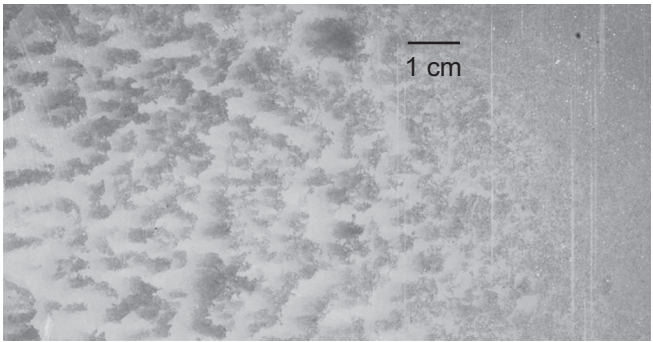}
\caption{Evidence for the influence of water vapor saturation on the laboratory-scale penitente instability. Dry air is injected from the left and progressively saturates in water vapor towards the right of the picture}
\label{Fig7}
\end{figure}

\subsection{Concluding remarks}
Interestingly, in both cases $\ell/\Lambda>1$ and $\ell/\Lambda<1$, the selected wavelength $\lambda_m \equiv 2\pi/k_m$ is found independent of the light penetration distance $\Lambda$. It is directly proportional to the distance $\ell$ from the ground at which the water vapor content does not feel the vapor flux modulation anymore. From the scaling law (\ref{kmsimple}), we obtain:
\begin{equation}
\lambda_m \sim 2\pi \frac{\mathcal{P}^{2/3}}{\Omega^{1/3}} \, \ell. 
\label{lambdamsimple}
\end{equation}
The factor of proportionality between $\lambda$ and $\ell$ is large, typically on the order of a few hundreds, for $\mathcal{P} \simeq 100$ and $\Omega$ in the range $0.1$--$0.5$. In the regime where the light penetration length $\Lambda$ is much larger than $\ell$ so that $1 \ll \mathcal{P} \ll \left( \Lambda/\ell \right)^2$, the result is similar with $\lambda_m \sim 2\pi \mathcal{P}^{2/3} \ell$.

This scaling law (\ref{lambdamsimple}) suggests that the emerging length-scale of penitentes is controlled by aerodynamic mixing above the ice surface. Molecular diffusion is inefficient compared to vapor advection. $\ell$ can therefore be interpreted as the distance to the soil at which mixing is efficient enough to recover a homogeneous vapor content. This length is set by the height over which turbulent fluctuations are suppressed close to the ground. In the field, $\lambda_m$, i.e. the peak separation of emerging penitentes, is typically on the order of a few tens of centimeters, which, according to (\ref{lambdamsimple}), corresponds to $\ell \simeq 0.1$~cm.  Assuming first that the ice surface is smooth, turbulent vapor mixing can hardly occur in the viscous sub-layer, whose thickness is $\ell_\nu \simeq 5 \nu/u_*$, where $\nu \simeq 10^{-5}$~m$^2$/s is the air kinematic viscosity and $u_*$ the wind shear velocity \cite{SG00}. $\ell=\ell_\nu$ would then correspond to $u_*$ of a few cm/s, i.e. low wind conditions, which is in agreement with observations that penitentes are specially developed on leeward slopes \cite{L54}. Moreover, thermal stratification may help to suppress turbulence even at larger winds. 

The ice surface is aerodynamically rough if the viscous length $\ell_\nu$ is smaller than the surface roughness $\delta$. $\ell = \delta$ on the order of a few mm is also reasonable. In the experiment \cite{BBB06}, micropenitentes emerge at $\lambda_m \simeq 1$~cm. A corresponding length $\ell$ on the order of $100~\mu$m is perfectly consistent with the surface roughness in these conditions.

Testing quantitatively the scaling law derived here requires better field data and/or experiments in which humidity is controlled in a precise way. Laboratory experiments such as \cite{BBB06} are performed in a confined environment and humidity in the experimental box is limited by a supply of dry air. As illustrated in Fig.~\ref{Fig7}, this usually establishes gradients along the air flow. It was accordingly reported in \cite{BBB06} that penitente emergence was eliminated when a moderate, steady breeze was induced to transport vapor. Further progress therefore requires an experimental control of hydrodynamic conditions. Building an experimental set-up able to control the mixing length $\ell$ is difficult. As the length-scales are larger, it may be easier to study the development of penitentes in the field, measuring the structure of the boundary layers in which vapor is transported away from the surface. 

Further theoretical progress requires a complete description of hydrodynamics. One expects a transition at large wind from penitentes to snow cups (or scallops), or towards `cross-hatching' features \cite{LM68,W71,M-GCC06}, similar in structure to regmaglypts. Interestingly, scallop patterns are also observed to form under dissolution, rather than ablation, processes \cite{BC74}, obeying similar scaling laws $\lambda_m \propto \nu/u_*$ \cite{T79}. As in Eq.~\ref{lambdamsimple}, the factor of proportionality is large, on the order of $10^3$ \cite{T79}. This suggests a common origin of the instability mechanism, where a Reynolds number is selected \cite{H81}, which must be investigated.

\noindent
\rule[0.1cm]{3cm}{1pt}

This work has benefited from the financial support of the Agence Nationale de la Recherche, `Zephyr' grant No. ERCS07\underline{\ }18.


\end{document}